# Overlapping Community Detection using Superior Seed Set Selection in Social Networks

Belfin R V[a], E.Grace Mary Kanaga[a], Piotr Bródka[b]

[a] Department of Computer Sciences Technology, Karunya Institute of Technology and Sciences, Coimbatore, India

[b] Department of Computational Intelligence, Faculty of Computer Science and Management, Wrocław University of Science and Management, Wrocław University of Science and Technology, Wrocław, Poland

ABSTRACT

Community discovery in the social network is one of the tremendously expanding areas which earn interest among researchers for past one decade. There are many already existing algorithms. However, new seed-based algorithms establish an emerging drift in this area. The basic idea behind these strategies is to identify exceptional nodes in the given network, called seeds, around which communities can be located. This paper proposes a blended strategy for locating suitable superior seed set by applying various centrality measures and using them to find overlapping communities. The examination of the algorithm has been performed regarding the goodness of the identified communities with the help of intra-cluster density and inter-cluster density. Finally, the runtime of the proposed algorithm has been compared with the existing community detection algorithms showing remarkable improvement.

## 1. INTRODUCTION

Some active social networks are very vibrant and add a huge amount of users at a particular interval. Each user in the social network will have their specifications and affinity in using certain products. Consider a company which wants its new products to be promoted to the customers through the social network. One way of promoting their products is to broadcast the product information to everyone in the network. This form of information propagation will be expensive and might be treated as spam by most of the social network users. Most of the victorious promotions in the social networks will be conducted by liking and sharing the content with the imminent ones in the network. The uncertainty arises when the marketing team wants to know "Where to start the promotion?". This situation prompts the researchers to work on an algorithm for finding the seed node from which the dispersion of information starts. For the larger network, there exists a necessity of having more seed nodes to proclaim the information faster across the whole network. This seed selection strategy also can be implemented in community detection to find out hidden communities available in the social network. The communities found will be crucial for the marketing team in a company to target marketing.

Heterogeneous networks are commonly utilized for modeling intercommunications in real-world systems in several areas, such as sociology, biology, knowledge spreading and transferring and numerous other areas(Han et al. 2016). One key topological property of real-world heterogeneous networks is that nodes are organized in tightly affiliated groups that are loosely connected to each other. Such groups are called as communities.Nodes forming a community are commonly agreed to share common proprieties and get involved in the same kind of functions(Kanawati 2014a).

There are many community detection methods proposed by researchers for the past few years. Disjoined community discovery is one of the biggest group of community detection method where every node in the network belongs only to one community(Fortunato 2010). The second group is overlapping community detection methods where one node can be a part of more than one community. Overlapping communities are valid in real social networks like Facebook, Twitter, and LinkedIn because the users will be involved in various ventures and associate with various communities (Xie, Kelley, and Szymanski 2013). Another aspect of the problem with overlapping community detection is the fuzzy nature of nodes concerning the degree of affinity towards each community in the network(Kundu and Pal 2015).

The basic idea of the strategy proposed in this paper is to find out a fraction of superior nodes of the input network, called superior seed set, around which local communities can be computed. The key concepts required to understand the proposed algorithm is given in the preliminaries section.

The rest of the paper is organized as follow. This chapter introduces the reader to basic concepts used in the paper. Chapter 2 presents state of the art in the field of community detection. In the 3$^{rd}$ chapter, the new approach is presented and evaluated in the 4$^{th}$ chapter. All findings are summarized in the last, 5$^{th}$ chapter.

*1.1 Preliminaries*

*1.1.1 Problem definition*

Given a graph $G = (V, E)$ with a set of nodes V and a set of edges E, we can denote the graph as an adjacency matrix A such that $A_{ij} = e_{ij}$ where $e_{ij}$ is the edge weight between vertices i and j, or $A_{ij} = 0$ if there is no edge. We can also determine various centrality measures $C_i$ for a node i to represent the strength of a node i in the entire graph. Let us assume the graph to be undirected. The conventional methods use any one of the centrality measures concerning the problem to define its seed set $S(G) = \{s_i, s_j, ..., s_\kappa\}$ where $\{s_i, s_j, ...s_k\} \in v$. Some algorithms pick seed set randomly which takes the parameter κ as input, to decide the number of seeds in the set. Every iteration the seeds will be adjusted to take the best possible position to cover the nodes around it. The difficulty in the conventional methods is that the seed set needs to be adjusted according to the circumstances of the problem. The proposed unified model determines the superior seed set S(G) from the centrality measures collectively. This article introduces a threshold value τ, which limits the number of seed nodes selected for the S(G).

*1.1.2 Centrality Measures*

Centrality measures will address the insights concerning a node's status in the whole social network. There are numerous alternatives of centrality measures possible with regard to the nature of the network. The commonly used centrality measures are the degree, betweenness, closeness, and the Eigenvector centrality.

- *Degree Centrality*

The degree is defined as the number of links it has with its neighbors. For an undirected graph, the degree will be the total number of links a node holds. Likewise, for a directed graph the nodes will own both "in-degrees" and "out-degrees". The in-degree of a node is the number of nodes incident on it, whereas, the out-degree of a node is defined as the number of links traveling apart from it.

Let A= $(a_{i,j})$ is being the adjacency matrix of a directed graph. The in-degree centrality $x_i$ of node i is given by:

$$x_i = \sum_k a_{k,i} \quad (1)$$

The out-degree centrality $y_i$ of node i is given by:

$$y_i = \sum_k a_{i,k} \quad (2)$$

- *Closeness*

Closeness is a measure which was first defined by Freeman in 1978. The real idea behind this measure is to identify the nodes which could influence other nodes in the network faster. The disadvantage of this measure is that it is not applicable to the graph containing separated components.

Suppose $d_{i,j}$ is the length of a geodesic path from i to j, meaning the number of edges along the path. Then the mean geodesic distance for vertex i is:

$$l_i = \frac{1}{n} \sum_j d_{i,j} \qquad (3)$$

The mean distance $l_i$ is not like other centrality measures because it gives the least value to the significant central nodes and vice versa. So, to determine closeness the equation (3) can be rewritten inversely as:

$$C_i = \frac{1}{l_i} = \frac{n}{\sum_j d_{i,j}} \qquad (4)$$

- *Betweenness*

Betweenness centrality is defined as the number of the shortest path from all the source nodes to every other target nodes that traverse through that node.

Mathematically, let $n^i_{s,t}$ be the number of geodesic paths from *s* to *t* that cross through *i* and let $n_{s,t}$ be the total number of geodesic paths from *s* to *t*. Where *s* is the source node and *t* is the target node. Then the betweenness centrality of the vertex *i* is:

$$b_i = \sum_{s,t} w^i_{s,t} = \sum_{s,t} \frac{n^i_{s,t}}{n_{s,t}} \qquad (5)$$

- *Eigenvector Centrality*

Eigenvector centrality is an extension of degree centrality. For instance, the in-degree measure provides one focal spot for each link a node receives. However, every vertex which has the same in-degree is not equal. Some nodes may be prominent than others. So, the eigenvector centrality determines the nodes are having high affinity with prominent nodes in the network as a high Eigenvector node.

Let $A=(a_{i,j})$ is being the adjacency matrix of a graph. The eigenvector centrality $e_i$ of node i is given by:

$$e_i = \frac{1}{\lambda}\sum_k a_{k,i} x_k \tag{6}$$

Where $\lambda \neq 0$ is a constant in matrix form.

$$\lambda x = xA \tag{7}$$

Therefore the centrality vector x is the left-hand eigenvector of the adjacency matrix A associated with the eigenvalue λ. It is smart to pick λ as the highest eigen value and an absolute value of the matrix A.

- *Page rank Centrality*

A node is powerful if it linked from other important and tightly linked nodes. There are three discrete factors that decide the Page rank of a node: (i) the number of links it receives, (ii) the link propensity of the nodes connected, and (iii) the centrality of the nodes connected.

$$p_i = \alpha \sum_j a_{ji} \frac{x_j}{L(j)} + \frac{1-\alpha}{N} \tag{8}$$

Where $L(j) = \sum_i a_{ji}$ is the number of neighbors of node $j$.

- *Local Clustering Coefficient*

The local clustering coefficient of a node in a network measures how close its neighbors is to become a complete graph.

The Local clustering coefficient for an undirected graph is given as

$$LCC_i = \frac{2|\{e_{jk} : v_j, v_k \in N_i, e_{jk} \in E\}|}{k_i(k_i - 1)} \tag{9}$$

Where $k_i$ is the number of neighbors of a vertex. The Neighborhood $N_i$ for a vertex $i$ is defined as it's immediately connected neighbors as follows:

$$N_i = \{v_j : e_{ij} \in E \lor e_{ij} \in E\} \qquad (10)$$

## 2. LITERATURE SURVEY

Seed selection strategy is one vital step in the influence propagation phenomenon(Weskida 2016)(Kanawati 2014a). But there are very limited modern works which show the seed selection strategy for community detection. These seeds then get expanded by some strategies to form a community structure. A recent survey by (Rossetti and Cazabet 2017) gives an unusual way of classifying community detection in social networks. A work by (Weskida 2016) showed a possibility of selecting the seeds by evolutionary algorithms. The authors have used the GPUs to compute the good seeds in the minimum amount of time. These methods using evolutionary algorithms should have a very strong and crisp evaluation function. These algorithms will fail in case of noisy evaluation functions. The work of (Yufeng Wang et al. 2014) selects the seed set by the page rank-like algorithm. The time aspect of selecting the seeds has been taken into consideration and simulated by (Sela and Pentland 2015).(Zhang et al. 2016) spoke about the intertwined influence propagation, which tells about the multiple product influence propagation strategies to the same seed. The research by (Rossetti et al. 2017) provides us with the community detection algorithm which tackles the dynamic nature of the social network. The work of (Yashen Wang et al. 2015)shows the co-ranking framework based seed selection for the heterogeneous network, did a comparative study of the seeding strategies in the complex social network.

Table 1: Characteristics of major seed centric algorithms

| Algorithm | Seed Nature | Seed Number | Seed Selection | Local Community |
|---|---|---|---|---|
| Licod (Kanawati 2011) | Set | Computed | Informed | Agglomerative |
| Yasca (Kanawati 2014b) | Single | Computed | Informed | Expansion |
| (Papadopoulos et al. 2012) | Subgraph | Computed | Informed | Expansion |
| (Whang, Gleich, and Dhillon 2013) | Single | Computed | Informed | Expansion |
| (Bollobas and Riordan 2009) | Subgraph | Computed | Informed | Expansion |
| (Whang, Gleich, and Dhillon 2016) | Set | Computed | Automatic | Expansion |

The community detection algorithms based on network decomposition which iteratively divides the network by removing links. The analysis performed in (Ding et al. 2016) is a network decomposition based community detection algorithm, finds the clique by using the greedy polynomial algorithm. Most of these algorithms select seed nodes randomly to do community discovery. Randomly selected seed nodes and communities require more iteration during optimization.

Label propagation based community detection algorithms take all the nodes as seeds initially. Then it iteratively adds nodes, which are similar to develop communities(Wu et al. 2016)(Zhao, Li, and Jin 2016).Recent work in (Zhi-Xiao et al. 2016) is based on node location analysis, which estimates the mass of the node. Mass of the node or seed rank can be estimated by the page rank centrality measure. Another recent work done in (Whang, Gleich, and Dhillon 2016) challenges that it determines the good seed node by applying the personalized page rank approach.

There are many researchers who claim that their approach of picking the seed node is the best way for the community discovery and information diffusion algorithms. However, the foremost motive behind the seed node is the ability of a node to influence more people and diffuse more information across the community and should have a close affinity with the members of the community

3. S‍UPERIOR S‍EED S‍ET S‍ELECTION ME‍THOD

The proposed model has two steps for community detection, (i) Superior Seed Set Selection (4-S) (ii) Superior Seed Set Expansion.

The 4-s model introduced here applies the fundamental centrality scores to detect superior seeds. The centrality scores are the key to express the strength of a node. Whatever problems we assume in social network analysis, the centrality scores will have a part in solving it. The proposed model uses the degree centrality, page rank centrality, local clustering coefficient and eigen vector centrality to find the superior seed set. The use of betweenness and closeness centrality will make the algorithm costly. The system utilizes set operators and a threshold value $\tau$ to limit the number of seeds. The value $\tau$ can be calculated by dividing the count of nodes in the graph with the split variable $\Delta$. The threshold $\tau$ can be given as,

$$\tau = \frac{n}{\Delta} \tag{11}$$

Where n is the total number of nodes in the graph and Δ is the split variable.

The principal concept behind the proposed methodology is to select the common superior nodes across all the centrality measures. The selected seeds S(g) will be superior in all centrality perspectives and can be used for any social network analysis problems. The number of seed nodes |S(g)| is equivalent to the number of communities in a community discovery problem.

*3.1 Superior Seed Set Selection (4-S)*

The proposed 4-S algorithm can be used with different centrality measures discussed in the preliminaries section. Figure 1 depicts the process of finding the superior seeds. This seed set can be used for various social network analysis problems.

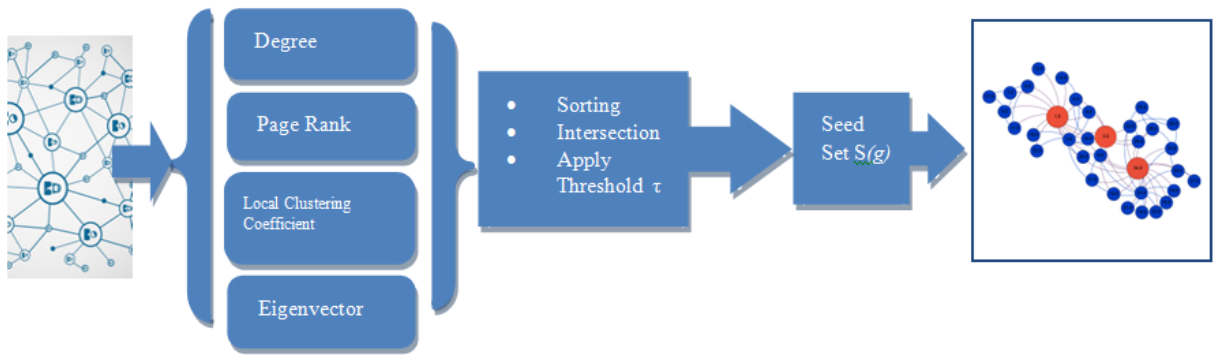

Figure 1: Proposed model for Superior seed set selection architecture

Here is the procedure for the superior seed selection algorithm:

a. We begin with a graph *G = (V, E)*, where *V* is the node set, and *E* denotes the edge set.

b. Compute the centrality measures degree d, local clustering coefficient l, eigenvector e, and page rank centrality p.

c. Define the centrality values as d = ( $d_i$, $d_j$,....,$d_n$), l = ($l_i$, $l_j$,....,$l_n$), e = ($e_i$, $e_j$,....,$e_n$) and p = ($p_i$, $p_j$,....,$p_n$). Where, n is the number of nodes in the graph.

d. Sort operation will be done with all the values of *d, l, e, p* in descending order. This operation will shuffle the nodes from superior to least superior order.

e. τ is the threshold value, which will be used to remove the insignificant nodes in the graph. Only the top τ nodes will be selected for the next step. The formula for calculating the τ value is specified in equation (11).

f. The sets $d_\tau$, $l_\tau$, $e_\tau$ and $c_\tau$ will be the input for the next step. Where, $d \supseteq d_\tau$, $l \supseteq l_\tau$, $e \supseteq e_\tau$ and $c \supseteq c_\tau$.

g. As a final step, the superior seeds can be determined by the intersection operation of sets is given by:

$$S(g) = d_\tau \cap l_\tau \cap e_\tau \cap c_\tau \qquad (12)$$

h. S(g) will hold the common nodes with higher rank from the centrality measures. |S(g)| is the count of the superior seed set.

---

**Algorithm 1:** Pseudo code of seed selection algorithm for selecting superior seed set.

---

***Input:*** *Objective Graph*

***Output:*** *Superior Seed Set*

*Read* graph G (V, E)

*Evaluate* degree centrality *d* of graph G (V, E)

*Evaluate* eigen value centrality *e* of graph G (V, E)

*Evaluate* local clustering coefficient *l* of graph G (V, E)

*Evaluate* page rank centrality *p* of graph G (V, E)

*Sort d, e, l, p*

***for*** split variable Δ ***do***

    Threshold τ = vertex count/*Δ*

    *Fetch* τ count of top nodes from list of d, e, l, p

    *Intersect* (d, e, l, p)

***end for***

---

*3.2 Superior Seed Set Expansion*

Seed expansion is a scheme of expanding the nodes from seed node to form a community. The expansion strategy will be different for various applications. The expansion method followed in this paper has been explained in Figure 2.

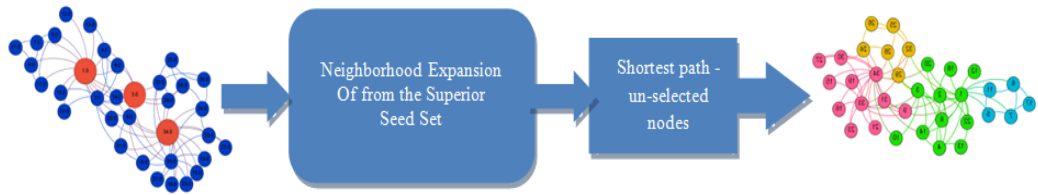

Figure 2: Superior Seed Set expansion method

The steps involved in seed expansion strategy are:

a. The Superior Seed Set S(g) will be determined from the 4-S Method.

b. A distance matrix $S(g)_{dist}$ will be prepared to find the distance between seeds.

c. The $\min(S(g)_{dist})$ will be estimated to define the expansion distance $Ex_{dist}$.

d. The Superior Seeds S(g) will be expanded to the $Ex_{dist}$ hops.

e. At this moment, An initial local community will be formed. Some unmapped nodes U(g) also will be comprised in the input network.

f. A distance matrix $U(g)_{dist}$ will be determined to calculate the distance between the seeds S(g) and the unmapped nodes U(g).

g. Each U(g) nodes will be added to the closest seed by referring the $U(g)_{dist}$.

h. Finally, all nodes from the input graph G will be added to various communities.

**Algorithm 2 :** Pseudo code of seed expansion algorithm for community discovery using superior seeds selected from the 4-S

| algorithm |
|---|
| *Input:* Seed set selected from 4-S Algorithm |
| *Output:* Overlapping community around each seed |
| Seed set *S(g)* Determination by 4-S Algorithm |
| Distance Matrix *S(g)* |
| Maximum Expansion Threshold (DistMatrix) *ExMAX$_\tau$* |
|     *for each(S (g))* **do** |
|         **While** (!ExMax$_\tau$) **do** |
|             Neighborhood expansion |
|         **end while** |
|     **end for each** |
| Determine Not in Community Nodes *Vφ* |
|     **for each** *(Vφ)* |
|         Neighbors of Vφ - $Nei_{V\varphi}$ |
|         Degree $Nei_{V\varphi}$ |
|         Assign node to the max degree nodes community |
|     **end for each** |

## 4. RESULT AND DISCUSSION

The proposed seed set expansion algorithm has been implemented in the real-world datasets and has given a better result. The internal and external density has been used in this work to prove the goodness of the cluster. Usually, for overlapping communities, internal and external metric values will be used. Because combination metrics and modularity scores will result in misleading values that should be inconsistent(Fortunato 2010). The goodness of the community using graph density ρ, intra-cluster density $\delta_{int}$ and inter cluster density $\delta_{ext}$ can be given as:

$$\delta_{int} > \rho > \delta_{ext} \tag{13}$$

Globally speaking, the internal density of a good clustering should be prominently greater than the density of the graph *g,* and the inter-cluster density of the clustering should be lower than the graph density.

*4.1 Inter-cluster and Intra-cluster density*

The cohesiveness of the edges in a graph can be readily attained by calculating the graph density ρ.

$$\rho = \frac{m}{n(n-1)/2}$$

(14)

Where, *n* is the count of nodes in the network, *m* is the count of edges and $n(n-1)/2$ is the maximum possible edges.

The density of the subgraph induced by the cluster as the internal or inter-cluster density $\delta_{int}$ (C) will be referred as:

$$\delta_{int}(C) = \frac{m_c}{n_c(n_c-1)/2}$$

(15)

The external or inter-cluster density $\delta_{ext}$(C) of clustering is defined as the ratio of inter-cluster edges to the highest amount of inter-cluster edges feasible.

$$\delta_{ext}(C) = \frac{m_c}{c(n-n_c)}$$

(16)

In equation (14) and (15), $m_c$ is the count of edges in the induced community, $n_c$ is the count of nodes in the community.

*4.2 Experiment*

The work has been tested with three real-world social networks. The dataset facts are given in Table 2. The step by step output for the Karate club dataset has been given, and the density based cluster goodness metric has been discussed for all the three datasets. Finally, the runtime of the proposed algorithm has been compared with some notable algorithms.

Table 2: Real-world networks used in the experiments

| Networks | Nodes | Edges | Average Degree | Description |
|---|---|---|---|---|
| Karate (Zachary 1977) | 34 | 78 | 4.588 | Zachary's Karate Club |
| Dolphin (Girvan and Newman 2002) | 62 | 159 | 5.129 | Dolphins Social Network |
| Football (Lusseau and Newman 2004) | 115 | 613 | 10.661 | American College Football |

The datasets are first treated with the 4-S algorithm to find out the superior seeds. Table 3 shows the assumed split variable $\Delta$, Threshold value $\tau$ and the length of the superior seed set $|S(g)|$ for the karate dataset. The $|S(g)|$ value gets smaller when the $\tau$ value increases. Figure 3 depicts the position of superior seed set in the karate club dataset.

Table 3: Superior Seed Set length with respect to the threshold for Karate Dataset

| $\Delta$ | 1 | 2 | 3 | 4 | 5 | 6 | 7 | 8 | 9 | 10 |
|---|---|---|---|---|---|---|---|---|---|---|
| $\tau$ | 34 | 17 | 11 | 8 | 7 | 6 | 5 | 4 | 4 | 3 |
| $|S(g)|$ | 34 | 10 | 6 | 5 | 4 | 3 | 3 | 2 | 2 | 2 |

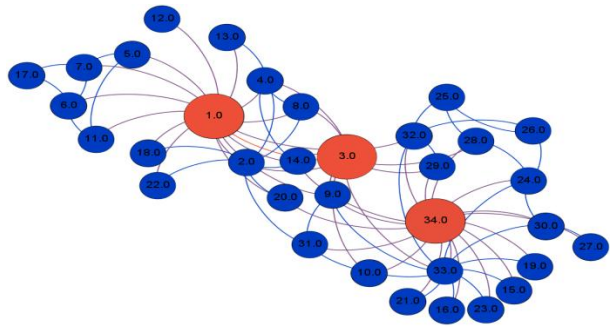

Figure 3: Superior Seed Set in karate club dataset with threshold 6 and 3 seeds

The selected seeds in Figure 3 will be expanded by the seed expansion algorithm given in the section superior seed set expansion to form a community structure given in Figure 4(b). The proposed algorithm not only finds out the overlapping community but also finds out the nested community in some cases. Figure 4 shows the communities detected for all the superior seed sets detected for Karate club dataset by the 4-S algorithm.

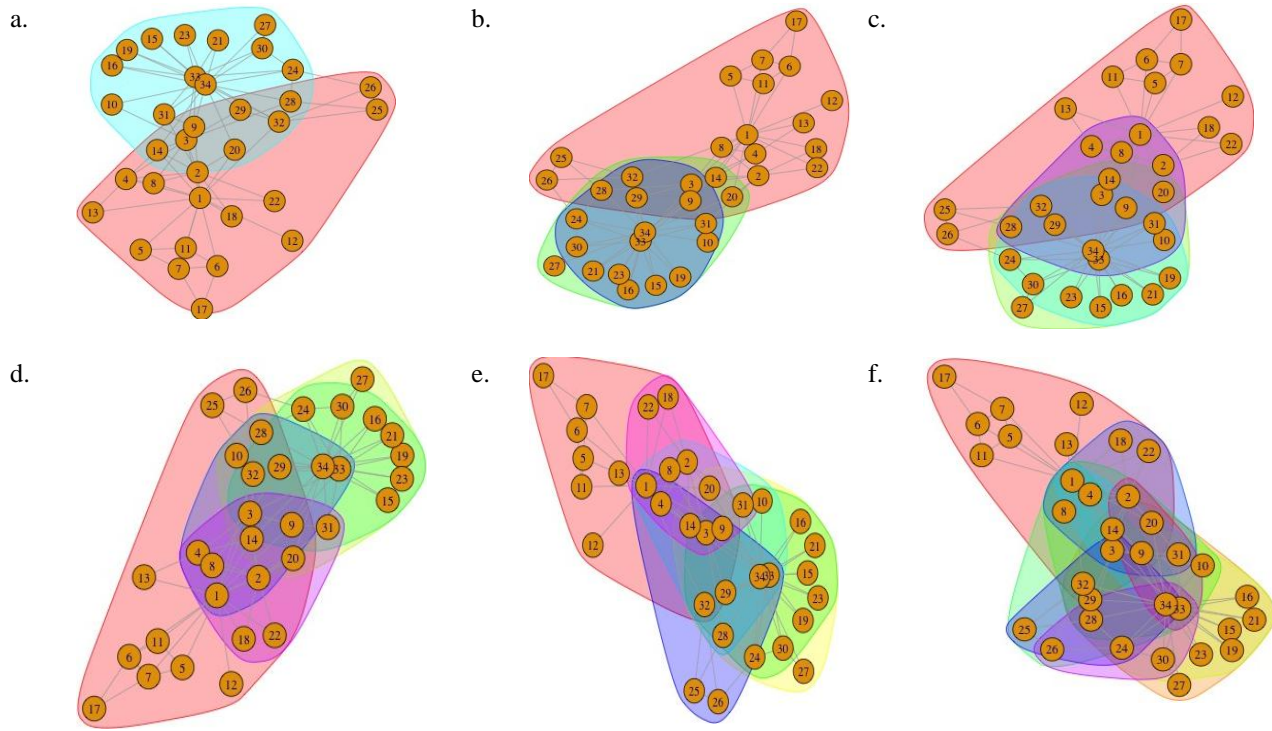

Figure 4: The figure a, b, c, d, e and f shows the community structure formation in Karate club dataset for the |S(g)| values 2, 3, 4, 5, 6 and 10 respectively

The communities formed by the 4-S Algorithm and the seed expansion strategy have been tested to identify its goodness. The $\delta_{int}$ (C) and $\delta_{ext}$ (C) has been determined for each community generated by the algorithm. Graph density $\rho$ for the input graph also should be calculated. The generated community will be good if it satisfies the equation (13). Figure 5 shows the line graph which compares the measure of densities for the karate data.

Similarly, the experiment has been done in dolphins and the football data set, and the results are shown in Figure 6 and Figure 7 respectively. All the datasets used go well with the equation 13. The plots prove that perfectly.

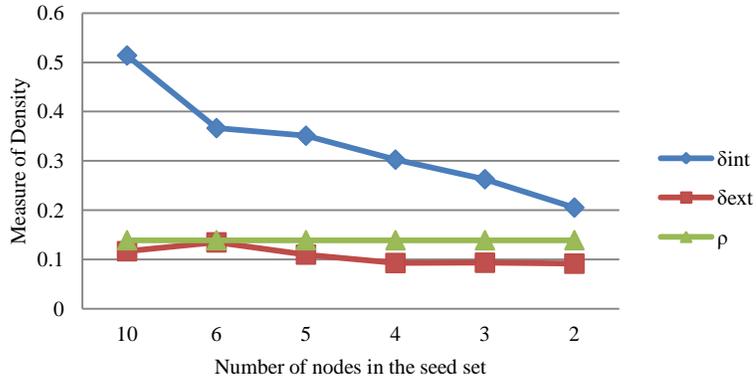

Figure 5: Comparison of intra-cluster, inter-cluster and graph density for karate dataset.

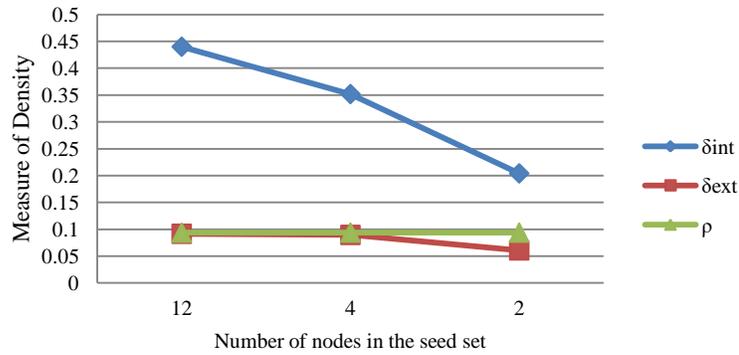

Figure 6: Comparison of intra-cluster, inter-cluster and graph density for dolphins dataset.

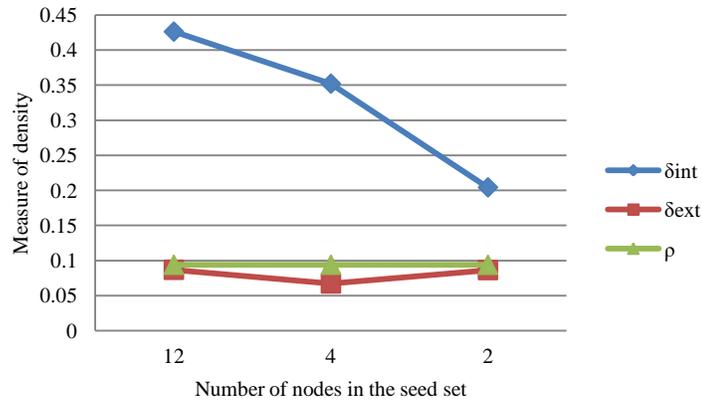

Figure 7: Comparison of intra-cluster, inter-cluster and graph density for football dataset.

Finally, the algorithm was compared with various community detection algorithms with respect to the runtime. Table 4 gives the detail of runtime of available algorithms Link Clustering (LC) (Ahn, Bagrow, and Lehmann 2010), Extended Link Clustering method (ELC) (Huang et al. 2013), Clique Percolation Method (CPM) (Palla et al. 2005), Overlapping

Cluster Generator (OCG) (Becker et al. 2012), Network Decomposition Overlapping Community Detection (NDOCD) (Ding et al. 2016) compared with the proposed algorithm. The runtime of the algorithm also shows good improvement for finding communities from the real-time datasets.

Table 4: Experiment result on three real-world networks

| Network | Runtime(s) | | | | | |
|---|---|---|---|---|---|---|
| | LC | ELC | CPM | OCG | NDOCD | 4-S |
| **Karate** | 0.61 | 2.49 | 0.67 | 0.23 | 0.2 | **0.18** |
| **Dolphin** | 1.97 | 8.96 | 2 | 0.25 | 0.2 | **0.19** |
| **Football** | 25.22 | 141.11 | 28.67 | 0.37 | 0.31 | **0.34** |

5. CONCLUSION

The proposed Superior Seed Set Selection and Seed Expansion strategies have been implemented and examined. The work conveys encouraging result and generates genuine communities. This algorithm generates excellent seeds automatically according to the threshold condition. The generated communities have proved that the selected seeds are excellent. The goodness of the community generated has been tested with the help of density based metric, which is a proven metric for finding the good communities. The runtime of the algorithm also gives a better improvement. When the proposed algorithm is compared with the fastest algorithm till now (Network Decomposition Overlapping Community Detection), it is 10% faster for the karate club dataset. Comparing runtime for dolphin dataset gives 5% improvement in speed. Unfortunately, the use of shortest path algorithm in the seed expansion model makes the process expensive for a graph with a big number of edges thus it needs to be replaced in our future work. We also need to evaluate other centrality measures, the range of threshold values and seed expansion approach in a future version of the Superior Seed Set Selection method.

ACKNOWLEDGEMENT

This work was partially supported by The National Science Centre, Poland, the project no. 2016/21/D/ST6/02408 and Wroclaw University of Science and Technology statutory funds.

**Belfin R V** is currently working as Assistant Professor in the Department of Computer Sciences Technology, Karunya Institute of Technology and Sciences. His research interests are in the area of Data Science specialized in Network science and ERP systems. Before his academic career, he was working as a technical consultant in a reputed software organization for over two years.

**E. Grace Mary Kanaga** is an Associate Professor in the Department of Computer Sciences Technology, Karunya Institute of Technology and Sciences. Her area of research includes Computational Intelligence, Big Data Analytics, and Brain-Computer Interface. She has published more than 43 papers in national and international conferences, international journals and book chapters.

**Piotr Bródka** is an assistant professor of Computer Science at Wroclaw University of Science and Technology. He was a visiting scholar at Stanford University in 2013. Piotr Bródka has authored over 70 research papers in the field of Network Science, on the extraction and dynamics of communities within social networks, spreading processes in complex networks and analysis of multilayer networks.